\documentclass{svproc}
\usepackage{url}
\usepackage{amsmath}
\usepackage{xcolor}
\usepackage{todonotes}

\begin{document}
\mainmatter              
\title{A new way to classify 2D higher order quantum superintegrable systems}
\titlerunning{Quantum superintegrable systems }  
%
\author{Bjorn K. Berntson\inst{1} \and Ian Marquette\inst{2}
 \and Willard Miller, Jr.\inst{3} }
\authorrunning{Bjorn K. Berntson et al.} 
%
\tocauthor{Ivar Ekeland, Roger Temam, Jeffrey Dean, David Grove,
Craig Chambers, Kim B. Bruce, and Elisa Bertino}
\institute{Department of Mathematics, KTH Royal Institute of Technology,
Stockholm, Sweden\\
\email{bbernts@kth.se}
\and
School of Mathematics and Physics, The University of Queensland,\\ Brisbane, Australia\\
\email{i.marquette@uq.edu.au},\\ WWW home page:
\texttt{https://smp.uq.edu.au/profile/211/ian-marquette}
\and
 School of Mathematics, University of Minnesota,\\con
Minneapolis, Minnesota, U.S.A.\\
\email{mille003@math.umn.edu},\\ WWW home page:
\texttt{http://www-users.math.umn.edu/~mille003}}

\maketitle              

\begin{abstract} 
We  revise a method by Kalnins, Kress and Miller (2010) for constructing a canonical form for symmetry operators of 
arbitrary order for the Schr\"odinger eigenvalue equation $H\Psi \equiv (\Delta_2 +V)\Psi=E\Psi$ on any 
2D Riemannian manifold, real or complex, that admits a separation of variables in some orthogonal 
coordinate system.  We apply the method, as an example,  to  revisit the Tremblay and Winternitz (2010) derivation of
 the Painlev\'e VI potential for a 3rd order superintegrable  flat space system that separates  in polar coordinates and, as new results,  we give a listing of the possible potentials on the 2-sphere that separate in spherical coordinates and all 2-hyperbolic (two-sheet) potentials separating in horocyclic coordinates. In particular, we  show that the Painlev\'e VI potential also appears for a 3rd order superintegrable system on the 2-sphere that separates in spherical coordinates, as well as a 3rd order superintegrable system on the 2-hyperboloid that separates in spherical coordinates and one that separates in horocyclic coordinates. Our aim is  to develop tools for analysis and classification of higher order superintegrable systems on any 2D Riemannian space, not just Euclidean space.
\end{abstract}
\section{Introduction}

In the paper \cite{KKM2010} the authors constructed a canonical form for symmetry operators of any order in 2D and used it to give the first  proof of the superintegrability  of the  quantum Tremblay, Turbiner,
 and Winternitz (TTW) system \cite{TTW} in polar coordinates, for all rational values of the parameter $k$. In the original method the various potentials were given and the problem was the construction of higher order symmetry operators that would verify superintegrability. The method was highly algebraic and required the solution of systems of difference equations on a lattice. Here, we consider   an arbitrary space admitting a separation in some orthogonal coordinate system (hence admitting a 2nd order symmetry operator), and search for all potentials $V$ for which the Schr\"odinger  equation admits an additional  independent symmetry operator of order higher than 2. Now the problem reduces to solving a system of partial differential equations. 
 
 In \S\ref{section2} we give a brief introduction to the method and then in \S\ref{section3} we specialize it to 3rd order superintegrable systems. More details can be found in \cite{KKM2010} and \cite{BMM2020}. In \S\ref{section4}   we treat a few important examples. We revisit the Tremblay and Winternitz derivation of
 the Painlev\'e VI potential for a 3rd order superintegrable  flat space system that separates  in polar coordinates, \cite{TW}, and we show among other new  results that the Painlev\'e VI potential also appears for a 3rd order superintegrable system on the 2-sphere that separates in spherical coordinates, as well as a 3rd order superintegrable system on the 2-hyperboloid that separates in horocyclic coordinates.

In \S\ref{section5} we classify all systems on the complex 2-sphere that admit a 3rd order symmetry operator and separate in spherical coordinates. For some of the cases where the system is 3rd order superintegrable we work out the symmetry algebra generated by the Hamiltonian $H$, the 2nd order symmetry operator $A$ associated with separation in spherical coordinates and the 3rd order symmetry operator $B$. In \S\ref{section6} we analyse the cases where the operators $A$ and $B$ are algebraically dependent and relate them to the Weierstrass $\wp$-function and its degenerations. In \S\ref{section7} we study superintegrable systems $\{H,A,B\}$ that are algebraically dependent but with $B$ of order higher than 3.  In \S\ref{section8} we classify 3rd order superintegrable systems on the two-sheet 2-hyperboloid that separate in
horocyclic coordinates. \S\ref{section9} is devoted to discussion.
Some of our principal results were announced in the proceedings paper \cite{BMM2020}, but here we give much more detail.

\section{The method} \label{section2}
We consider a Schr\"odinger equation on a 2D real or complex Riemannian manifold with Laplace-Beltrami operator $\Delta_2$ and potential $V$:
\begin{equation} \label{TIS}H\Psi\equiv (-\frac{\hbar^2}{2}\Delta_2+V)\Psi=E\Psi \end{equation}
that also  admits an orthogonal  separation of variables. 
If $\{u_1,u_2\}$ is the  orthogonal separable coordinate system   the corresponding Schr\"odinger
operator can always be put in the  form 
\begin{equation} \label{canform}
H=-\frac{\hbar^2}{2}
\Delta_2+V(u_1,u_2)=\end{equation}
\[\frac{1}{f_1(u_1)+f_2(u_2)}\left(-\frac{\hbar^2}{2}\partial^2_{u_1}-\frac{\hbar^2}{2}\partial^2_{u_2}+V_1(u_1)+V_2(u_2)
\right)
\]
and, due to the separability, there is the second-order symmetry operator
\begin{equation} \label{Lequation}
L= \frac{f_2(u_2)}{f_1(u_1)+f_2(u_2)}\left(-\frac{\hbar^2}{2}\partial^2_{u_1}+V_1(u_1)\right)\end{equation}
\[-\frac{f_1(u_1)}{f_1(u_1)+f_2(u_2)}\left(-\frac{\hbar^2}{2}
\partial^2_{u_2}+V_2(u_2)\right),
\]
i.e., $
[H,L]=0$.
We  look for a  partial differential symmetry   operator of arbitrary order  ${\tilde L}(H,L,u_1,u_2)$ that satisfies 
\begin{equation}
[H,{\tilde L}]= 0.\label{newconditions1}
\end{equation}
We require that the symmetry operator take the standard form
\begin{align}\label{standardLform}
{\tilde
L}= \sum_{j,k} &\left( A^{j,k}(u_1,u_2)\partial_{u_1u_2}-B^{j,k}(u_1,u_2)\partial_{u_1}
-C^{j,k}(u_1,u_2)\partial_{u_2}\right. \\
&\left. + D^{j,k}(u_1,u_2) \nonumber
\right)H^jL^k.
\end{align}

This can always be done. More details of the derivation can be found in \cite{KKM2010} and \cite{BMM2020}.

In this view we can write
\begin{equation}\label{generalLform}
{\tilde
L}(H,L_2,u_1,u_2)=A(u_1,u_2)\partial_{u_1 u_2}-B(u_1,u_2)\partial_{u_1}
-C(u_1,u_2)\partial_{u_2}
+ D(u_1,u_2),
\end{equation}
and consider $\tilde  L$ as an at most second-order  order differential operator in $u_1,u_2$ that is analytic in the parameters $H,L$. 
Then the above system of equations can be written in the more compact form {\small
\begin{equation} \label{partialxyHL}
\partial_{u_1}^2A+\partial_{u_2}^2A-2\partial_{u_2}B-2\partial_{u_1}C
=0,
\end{equation}
\begin{equation} \label{partialxHL}
\frac{\hbar^2}{2}(\partial_{u_1}^2B+\partial_{u_2}^2B)-2\partial_{u_2}A\,V_2-\hbar^2\partial_{u_1}D-AV'_2+(2\partial_{u_2}A\,f_2+Af'_2)H-2\partial_{u_2}A\,L_2=0,
\end{equation}
\begin{equation} \label{partialyHL}
\frac{\hbar^2}{2}(\partial_{u_1}^2C+\partial_{u_2}^2C)-2\partial_{u_1}AV_1-\hbar^2\partial_{u_2}D-AV'_1+(2\partial_{u_1}A\,f_1+Af'_1)H+2\partial_{u_1}A\,L_2=0,
\end{equation}
\begin{equation}\label{constanttermHL}
-\frac{\hbar^2}{2}(\partial_{u_1}^2D+\partial_{u_2}^2D)+2\partial_{u_1}B\,V_1+2\partial_{u_2}C\,V_2+BV'_1+Cv'_2
\end{equation}
\[
-(2\partial_{u_1}B\,f_1+2\partial_{u_2}C\,f_2+Bf'_1
+Cf'_2)H+(-2\partial_{u_1}B+2\partial_{u_2}C)\,L_2=0.
\] }

We can view (\ref{partialxyHL}) as an equation for $A,B,C$ and (\ref{partialxHL}), (\ref{partialyHL}) as the defining equations for $\partial_{u_1}D, \partial_{u_2}D$.
Then $\tilde L$ is $\hat L$ with the terms in $H$ and $L$ interpreted as (\ref{standardLform}) and considered as partial differential operators.

We can simplify this system by noting that there are two functions\hfill\break $F(u_1,u_2,H,L)$, $G(u_1,u_2,H,L)$ such
that (\ref{partialxyHL}) is satisfied by 
\begin{equation}\label{ABCtoFG}
A=F,\qquad B=\frac12 \partial_{u_2}F+\partial_{u_1}G,\qquad C=\frac12\partial_{u_1} F-\partial_{u_2}G.
\end{equation}
Then the integrability condition for (\ref{partialxHL}), (\ref{partialyHL}) is (with the shorthand notation $\partial_{u_j}F=F_j$, $\partial_{u_j}\partial_{u_\ell}F=F_{j\ell}$, etc., for $F$ and $G$),

{\small
\begin{align}&\hbar^2 G_{1222}+\frac14 \hbar^2 F_{2222}-2F_{22}(V_2-f_2H+L_2)
-3F_{2}(V'_2-f_2'H)-F(V''_2-f''_2H) \label{eqn1}\\
 &=-\hbar^2 G_{1112} +\frac14 \hbar^2 F_{1111}-2F_{11}(V_1-f_1H-L_2)
-3F_{1}(V'_1-f'_1H)-F(V''_1-f''_1H),\nonumber \end{align} }
and equation (\ref{constanttermHL}) becomes  {\small
\begin{align}
&\frac14 \hbar^2 F_{1112}-2F_{12}(V_1-f_1H)-F_{1}(V'_2-f'_2H)+\frac14 \hbar^2
G_{1111} \label{eqn2}\\
&-2G_{11}(V_1-f_1H-L_2)-G_{1}(V'_1-f'_1H)=-\frac14\hbar^2F_{1222}+2F_{12}(V_2-f_2H) \nonumber \\
&+F_{2}(V'_1-f'_1H)+\frac14 \hbar^2
G_{2222}-
2G_{22}(V_2-f_2H+L_2)-G_{2}(V'_2-f'_2H).\nonumber\end{align}
We remark that any solution of (\ref{eqn1}), (\ref{eqn2}) with $A,B,C$ not identically $0$ corresponds to a symmetry operator that does not commute with $L$, hence is algebraically  independent of the symmetries $H, L$.

\section{3rd order superintegrability}\label{section3}
To show how equations (\ref{eqn1}) and (\ref{eqn2}) can be used to find potentials for superintegrable systems, we  provide detailed derivations of the determining equations for 3rd order superintegrability.  First we note that the most general 3rd order operator must be of the form (\ref{standardLform}) with  {\small
\begin{eqnarray*} &A&=A^0(x,y),\quad B=B^0(x,y)+B^H(x,y) H+B^L(x,y)L,\\
&C&=C^0(x,y)+C^H(x,y) H+C^L(x,y)L,\
 D=D^0(x,y)+D^H(x,y) H+D^L(x,y)L,\end{eqnarray*} }
 or, in view of (\ref{ABCtoFG}),
\begin{equation}\label{eqn3} F(x,y)=F^0(x,y),\quad G(x,y)=G^0(x,y)+G^H(x,y)H +G^L(x,y)L.\end{equation}

Substituting (\ref{eqn3}) into (\ref{eqn1}), (\ref{eqn2}) and noting that the coefficients of independent powers of $H$ and $L$ in these expressions must vanish, we obtain 9 equations, (the first 3 from (\ref{eqn1}) and the next 6 from (\ref{eqn2})):

\begin{eqnarray*}&0=&-6V_1'F^0_1+6V_2'F^0_2-4V_1F^0_{11}+4V_2F^0_{22}-2\hbar^2G^0_{1112}-2\hbar^2G^0_{1222}\\
&&+2F^0V_2''-2F^0V_1'',\\
&0=& F^0_{11}+F^0_{22},\\
&0=&-\hbar^2G^H_{1112}-\hbar^2G^H_{1222}+3f_1'F^0_1-3f_2'F^0_2+2f_1F^0_{11}-2f_2F^0_{22}-F^0f_2''+F^0f_1'',\\
&0=&V_2'F^0_1+V_1'F^0_2+V_1'G^0_1-V_2'G^0_2+2F^0_{12}V_2+2F^0_{12}V_1+2V_1G^0_{11}-2V_2G^0_{22}-\\
&&\frac14\hbar^2G^0_{1111}+\frac14\hbar^2G^0_{2222},\\ 
&0=&V_1'G^L_1-V_2'G^L_2+2V_1G^L_{11}-2G^0_{11}-2V_2G^L_{22}-2G^0_{22},\\
&0=& G^L_{11}+G^L_{22},\\
&0=& -f_2'F^0_1-f_1'F^0_2+V_1'G^H_1-f_1'G^0_1-V_2'G^H_2+f_2'G^0_2-2F^0_{12}f_2-2F^0_{12}f_1+2V_1G^H_{11}\\
&&-2f_1G^0_{11}-2V_2G^H_{22}+2f_2G^0_{22}-\frac14\hbar^2G^H_{1111}+\frac14\hbar^2G^H_{2222},\\
&0=&-f_1'G^L_{1}+f_2'G^L_{2}+2f_2G^L_{22}-2f_1G^L_{11}-2G^H_{11}-2G^H_{22},\\
&0=&-f_1'G^H_1+f_2'G^H_2+2f_2G^H_{22}-2f_1G^H_{11}.\\
\end{eqnarray*}

\section{Some examples (mostly new)}\label{section4} For our first examples we are particularly interested in potentials with nonlinear defining equations. First, we show that we can obtain the result of Tremblay and Winternitz \cite{TW} that the  quantum system separating in polar coordinates in 2D Euclidean space admits potentials that are expressed in terms of the sixth Painlev\'e transcendent or in terms of the Weierstrass $\wp$-function. To do this we must put the system in the canonical form (\ref{canform}). The separable polar coordinates are $(x,y)=(r\cos(\theta),r\sin(\theta))$. For the canonical form we have $r=\exp(u_1),\ \theta =u_2$. Thus, $f_1(u_1)=\exp(2u_1)$ and $f_2(u_2)=0$. We know that these Painlev\'e VI can appear only if the potential depends  on the angular variable alone, so we set $V_1(u_1)=0$.  Since we want only systems that satisfy nonlinear equations alone, whenever an explicit linear equation for the potential appears, we require that it vanish identically. We have the freedom to replace the angular variable $u_2$ by $u_2+c$ for some real constant $c$ to simplify the expressions. Also we can rescale the answer.

We obtain a solution of the equations in \S\ref{section3} as 
\begin{eqnarray*} F^0 &=& 4\hbar^2\exp(-u_1)\sin(u_2),\quad G^L= -8 \exp(-u_1)\cos(u_2)+a_4 u_2+a_3,\\  G^0 &=& -U_1(u_2)\exp(-u_1)+U_2(u_2),\quad
 G^H=a_5,\end{eqnarray*}
subject to the conditions
\begin{eqnarray}\label{cond1}&0=&a_4\frac{dV_2}{du_2}+2\frac{d^2U_2}{du_2^2},\\ \label{cond2}&0=& \hbar^2\frac{d^4U_2}{du_2^4}+4a_4\frac{dV_2}{du_2}V_2-4\frac{dV_2}{du_2}\frac{dU_2}{du_2},\\
\label{cond3}&0=&8V_2\cos(u_2)+4\frac{dV_2}{du_2}\sin(u_2)-\frac{d^2U_1}{du_2^2}- U_1,\\
\label{cond4}&0=&\frac{dV_2}{du_2}\frac{dU_1}{du_2}-\hbar^2\frac{d^3V_2}{du_2^3}\sin(u_2)-4\hbar^2 \frac{d^2V_2}{du_2^2}\cos(u_2)\\
&&+2\sin(u_2)(\hbar^2+4V_2)\frac{dV_2}{du_2}+2V_2\left(6\hbar^2\cos(u_2)+8V_2\cos(u_2)-U_1\right) \nonumber
.\end{eqnarray}

There are basically two cases to consider: \begin{enumerate} \item $a_4=0$. \\ \medskip

Then condition (\ref{cond1}) says that $U_2$ is linear in $u_2$. Thus condition (\ref{cond2}) is a linear equation for $V_2(y)$ which must vanish. Then condition (\ref{cond3}) can be solved for $U_1(y)$ and the result substituted into condition (\ref{cond4}) to obtain an equation for $V_2(u_2)$. After some manipulation (using the fact that $V_2$ is unchanged by transformations $W\to W+c$, where $c$ is a constant), we obtain an equation characterizing Painlev\'e VI, in agreement with \cite{TW}, equation (4.27):
\begin{equation} \label{PVI} \hbar^2\left(\sin(u_2) \frac{d^4W}{du_2^4}+4\cos(u_2)\frac{d^3W}{du_2^3}-6\sin(u_2)\frac{d^2W}{du_2^2}-4\cos(u_2)\frac{dW}{du_2}\right)   \end{equation}
\[ -12\sin(u_2)\frac{dW}{du_2}\frac{d^2W}{du_2^2}-4\cos(u_2)W\frac{d^2W}{du_2^2}-4(\beta_1\sin(u_2)-\beta_2\cos(u_2))\frac{d^2W}{du_2^2}\]
\[-16\cos(u_2)(\frac{dW}{du_2})^2+8\sin(u_2)W\frac{dW}{du_2}-8(\beta_1\cos(u_2)+\beta_2\sin(u_2))\frac{dW}{du_2}=0\] Here $V_2(u_2)=\frac{dW(u_2)}{du_2}$.
\item $a_4 \ne 0$. \medskip\\   Solving condition (\ref{cond1}) for $V_2(u_2)$ and substituting the result and (\ref{cond1}) into (\ref{cond2}) we obtain the equation that characterizes the Weierstrass $\wp$-function (in fact it is a translated and rescaled version):
\begin{equation} \label{Weierstrass} \hbar^2 \frac{d^3V_2}{du_2^3}-12\frac{dV_2}{du_2}V_2+12a_1\frac{dV_2}{du_2}=0,\end{equation} 
where $a_1$ is an arbitrary constant. Thus $V_2(u_2)=\hbar^2\wp(u_2-u_{2,0};g_2,g_3)+a_1$, where $u_{2,0}$, $g_2$, and $g_3$ are arbitrary constants. As shown in \cite{TW} this solution is subject to the compatibility condition (\ref{cond3}) and (\ref{cond4}), which leads to a complicated nonlinear differential equation for $V_2(u_2)$.
\end{enumerate}

Now we consider the analogous system on the 2-sphere, separable in spherical coordinates. Here $s_1=\sin(\theta)\cos(\phi),\ s_2=\sin(\theta)\sin(\phi),s_3=\cos(\theta)$ with $s_1^2+s_2^2+s_3^2=1$. This system is in canonical form with coordinates $\{u_1,u_2\}$ where
\begin{equation}\label{cansp} \sin(\theta)=(\cosh(u_1))^{-1},\ \phi=u_2,\ f_1(u_1)=(\cosh(u_1))^{-2},\ f_2(u_2)=0. \end{equation} As before we look for solutions such that $V_1(u_1)=0$ and $V_2(u_2)$ satisfies a nonlinear equation only.

The computation is very similar to that for the Euclidean space example. We obtain the solution 
\begin{equation} \label{nonlinearsoln} F^0 = 4\hbar^2\cosh(u_1)\sin(u_2),\ G^L= 8\sinh(u_1)\cos(u_2)+a_{4}y+a_3,\end{equation}
\[G^0 =  \sinh(u_1)\ U_1(u_2)+U_2(u_2),\
 G^H=a_5,\]
 subject to the conditions (\ref{cond1}-\ref{cond4}), exactly the same as for Euclidean space. 
  Thus the system on the 2-sphere also admits Painlev\'e VI and special Weierstrass $\wp$-function potentials for 3rd order superintegrability. It is clear from these results that these systems in Euclidean space can be obtained as B\^ocher contractions, \cite[Chapter 15]{Bocher}, of the corresponding systems on the 2-sphere. 

\bigskip

Next we consider spherical coordinates  on the hyperboloid $s_1^2-s_2^2-s_3^2=1$,
\[ s_1=\cosh(x),\ s_2=\sinh(x)\cos(\phi),\ s_3=\sinh(x)\sin(\phi).\]
For the canonical form we find  
\[ \tanh \bigg(\frac{u_1}{2}\bigg)=\exp(x),\ u_2=\phi,\quad  f_1(u_1)=\frac{1}{\sinh^2(u_1)},\ f_2(u_2)=0,\]
 and we look for solutions such that $V_1(u_1)=0$ and $V_2(u_2)$ satisfies only a nonlinear equation. We obtain the solution 
\[ F^0 = 4\hbar^2\sin(u_2)\sinh(u_1),\ G^L= 8\cos(u_2)\cosh(u_1)+a_{4}u_2+a_3,\]
\[G^0 =  \cosh(u_1)\ U_1(u_2)+U_2(u_2),\
 G^H=a_5,\]
 subject to the conditions (\ref{cond1})-(\ref{cond4}), again exactly the same as for flat space. Thus the system on the 2-hyperboloid  admits Painlev\'e VI and special Weierstrass $\wp$-function potentials for 3rd order superintegrability.
 
For our next example we consider horocyclic coordinates $\{u_1,u_2\}$ on the hyperboloid $s_1^2-s_2^2-s_3^2=1$, e.g. \cite[Section 7.7]{KKM2018}:
 \begin{equation}\label{horcoords} s_1=\frac12\bigg(u_1+\frac{u_2^2+1}{u_1}\bigg), \ s_2=\frac12\bigg(u_1+\frac{u_2^2-1}{u_1}\bigg), \ s_3=\frac{u_2}{u_1}.\end{equation}
These coordinates are separable and the canonical system is defined by $f_1(u_1)=1/u_1^2$, $f_1(u_2)=0$. We look for systems such that $V_1(u_1)=0$, in analogy with our first three examples.
 
We obtain the solution 
\begin{eqnarray*}F^0 &=&- \frac12 a_8 \hbar^2u_1,\quad G^L= \frac{u_1^2(a_8 u_2+a_9)}{2}-\frac{a_8u_2^3}{6}-\frac{a_9u_2^2}{2}+a_{10}u_2,\\
G^0& =& \frac{u_1^2}{2}U_1(u_2)+U_2(u_2),\quad
 G^H=a_7,\end{eqnarray*}
 subject to the conditions 
 \begin{eqnarray}&0=& a_8\frac{dV_2}{du_2}+2\frac{d^2U_1}{du_2^2},\label{hcond1}\\
 &0=& \frac12\hbar^2 a_8\frac{d^3V_2}{du_2^3}-4a_8\frac{dV_2}{du_2}V_2+4\frac{dV_2}{du_2}\frac{dU_1}{du_2},\label{hcond2}\\
  &0=&(2a_{10}-2a_9 u_2 -a_8u_2^2)\frac{dV_2}{du_2}-4(a_9+a_8 u_2)V_2+4U_1+4\frac{d^2U_2}{du_2^2},\label{hcond3}\\
 &0=&   -2\hbar^2 a_8 u_2^2\frac{dV_2}{du_2}+16(a_9+a_8 u_2)V_2^2-4(2a_{10}-2a_9u_2+a_8 u_2^2)\frac{dV_2}{du_2}V_2 \label{hcond4} \\
   &&\quad +\frac{\hbar^2}{2}(2a_{10}-2a_9 u_2-a_8u_2^2)\frac{d^3V_2}{du_2^3}-4\hbar^2(a_9+a_8 u_2)   \nonumber \\
   &&\quad -16V_2U_1+8\frac{dV_2}{du_2}\frac{dU_2}{du_2}.   \nonumber \end{eqnarray}
   There are again two basic cases here: \begin{enumerate} \item $a_8=0$. \\ 

Then conditions (\ref{hcond1}) and (\ref{hcond2}) say that $U_1$ is a constant: $U_1(u_2)=d_1$.  Then condition (\ref{hcond3}) can be solved for $U_2(u_2)$ and the result substituted into condition (\ref{hcond4}) to obtain an equation for $V_2(u_2)$:
\begin{equation}\label{horocyclicnonlinear}   -4a_9(\frac{dW}{du_2})^2+\left((-3a_9u_2+3a_{10})\frac{d^2W}{du_2^2} +4d_1\right)\frac{dW}{du_2}+\end{equation}
\[(-a_9 W+2d_1u_2-2d_3)\frac{d^2W}{du_2^2}
+\hbar^2a_9\frac{d^3W}{du_2^3}-\frac14\hbar^2(a_{10}-a_9u_2)\frac{d^4W}{du_2^4}=0,\]  
where $ V_2(u_2)=\frac{dW(u_2)}{du_2}$.  

With the integrating factor $\mu(u_2)=u_2 a_{9}-a_{10}$, the fourth order nonlinear differential equation admits the following first integral
\begin{eqnarray} J&=&\frac{1}{4}\hbar^{2}( a_{10}-a_{9} u_2)^{2} \frac{d^{3}W}{du_2^{3}} -\frac{\hbar^2}2(a_{10}a_{9}\hbar^{2}-a_{9}^{2}\hbar^{2})\frac{d^{2}W}{du_2^{2}} \\
&&-\frac32 ( a_{10}^{2} -2 a_{10}a_{9} u_2- a_{9}^{2} u_2^{2}) \bigg(\frac{dW}{du_2}\bigg)^{2}+  ( a_{10}a_{9}-a_{9}^{2} u_2) W \frac{dW}{du_2}\nonumber \\
&&+( 2 a_{10} d_{3} -\frac{a_{9}^{2} \hbar^{2}}{2} - 2 a_{10} d_{1} u_2 - 2 a_{9} d_{3} u_2 + 2 a_{9} d_{1} u_2^{2} ) \frac{dW}{du} \nonumber\\
&&+\frac{1}{2}a_{9}^{2}W^{2}+ 2 a_{9} d_{3} W - 2 a_{10} d_{1}. \nonumber \end{eqnarray}

We now consider two subcases.

\begin{enumerate}

\item $a_9\neq 0$. Using the following transformation of the independent and dependent variables
\begin{equation}
W(u_2) =    u_2  \tilde{W}(u_2) + b + c u_2^2 + d u_2^4, \quad z = -\frac{a_{10}}{a_{9}} + u
\end{equation}
and further using
\begin{equation}y = z^{2} \end{equation}
and using the constraint $d_3 = \frac{a_{10}d_{1}}{a_{9}}$ (for $a_{9} \neq 0$), we obtain a third order differential equation can be related to the Chazy I equation. This equation appears in the classification, \cite[Eq. A.3]{Cosgrove2000b}. It has the form
\[  \tilde{W}''' =-\frac{2}{f^{2}(y)} ( 3 c_{1} y ( y \tilde{W}' - \tilde{W})^{2} + c_{2} ( y \tilde{W}' -\tilde{W}) ( 3 y \tilde{W}'-\tilde{W})\]
\[+ c_{3} \tilde{W}' ( 3 y \tilde{W}' -2 \tilde{W}) + 3 c_{4} (\tilde{W}')^{2} + 2c_{5} y (y \tilde{W}'-\tilde{W}) \]
\[ + c_{6} ( 2 y \tilde{W}'-\tilde{W}) + 2c_{7} \tilde{W}' + c_{8} y +c_{9})- \frac{f'}{f} \tilde{W}'', \]
where
\[ f(y)= c_{1} y^{3} +c_{2} y^{2} +c_{3} y +c_{4}\]
Our case corresponds to the following choice of parameters:
\begin{eqnarray*}
 &&c_{1} = c_{2} =0 , \quad c_{3} = 2 a_{9}, \quad c_{4}= c_{5} = 0, \quad c_{6} = -2 (2 a_{9} c + d_{1}), \\
  &&c_{7} = \frac{1}{4} a_{9} (8 b - a_{9}),\quad  c_{8} = 2 (a_{9} c^{2} + c d_{1}),   \\
 &&c_9=\frac{1}{2} (-8 a_{9} b c - 4 b d_{1} - a_{9}^{2} c),\quad 
 d = 0, \quad a_{9}= - \hbar^{2}.
 \end{eqnarray*}

This equation can be further integrated, \cite[Eq. A.21]{Cosgrove2000b} and the resulting equation takes the following form
\begin{eqnarray}
 (\tilde{W}'')^{2}&=&-\frac{4}{f^{2}}\big( c_{1} ( x \tilde{W}'- \tilde{W})^{2}+ c_{2} \tilde{W}' ( y \tilde{W}'-\tilde{W})^{2}+c_{3} (\tilde{W}')^{2} (y \tilde{W}'-\tilde{W}) \nonumber\\
 && + c_{4} (\tilde{W}')^{3} +c_{5}(y \tilde{W}' -\tilde{W})^{2}+ c_{6}\tilde{W}'(y\tilde{W}'-\tilde{W}) +c_{7} (\tilde{W}')^{2} \nonumber \\
 &&+c_{8} (y\tilde{W}'-\tilde{W})+c_{9}\tilde{W}' +c_{10}\big)
 \end{eqnarray}
where $c_{10}$ is an integration constant. This equation is known as SD-I and was discussed in \cite[Eq. 4.9]{Cosgrove1993}. It has 6 subcases: SD-Ia, SD-Ib, SD-Ic, SD-Id, SD-Ie, SD-If.
Here the non zero parameters are $c_{3}$, $c_{6}$, $c_{6}$, $c_{8}$ , $c_{9}$, according to \cite{Cosgrove2000b}. As described in \cite[Eq. 5.5]{Cosgrove1993} the equation is related to Painlev\'{e} III and Painlev\'{e} V equations. Explicit formulas are given in \cite[Eqs. 5.26-5.40]{Cosgrove1993}.

\item $a_{9}=0$. We obtain the third order equation
\[-2 d_{1} W + ( 2  d_{3} - 2  d_{1} u_2) W'
-\frac{3}{2} a_{10} W'^{2} + \frac{1}{4} a_{10} \hbar^{2} W'''=0,\]
which is a special case of \cite[Eq. 83]{MSW}:
\begin{eqnarray}
&& \alpha \hbar^{2} W''' - 6 \alpha W'^{2} - 4 ( c_{1} u^{3} - \alpha c_{2} u^{2}+b_{1} u +b_{0} )W' \nonumber\\
&&- 4 ( 3 c_{1} u^{2}- 2 c_{2} \alpha u +b_{1})W  -\frac{2}{3} c_{2}^{2} \alpha u^{4} 
 + 4 ( \frac{1}{3} c_{2} b_{1} -c_{1}a_{2})u^{3} \nonumber \\
 &&-2( 3 c_{1} a_{1} -2 c_{2}b_{0})u^{2} + k_{2}u +k_{4}=0.
\end{eqnarray}
This equation admits the first integral 
\begin{eqnarray}
J=&&\alpha \hbar^{2} W' -\alpha W^{2} - ( \alpha a_{1} - 2(b_{0}-\alpha a_{2}) u - 2 b_{1} u^{2} -\frac{2}{3} \alpha c_{2} u^{3}- 2c_{1}u^{4})W \nonumber\\
&& -\frac{1}{9} \alpha c_{2}^{2} u^{6} + \frac{1}{6} ( 3 a_{2} c_{1} -b_{1}c_{2}) u^{5} + (\frac{2}{3}(\alpha a_{2}c_{2}-b_{0}c_{2}) +a_{1}c_{1})u^{4} \nonumber \\
&& + (\frac{4}{3}\alpha a_{1} c_{2}-\frac{k_{2}}{4})u^{3}-\frac{1}{8}(\alpha k_{1} +4 k_{4})u^{2} +k_{5}u -\alpha \frac{k_{3}}{2} =0. 
\end{eqnarray}
Then, a Cole-Hopf transformation $W=-\hbar^{2} \frac{U'}{U}$ gives a second order linear equation for $U$.

\end{enumerate}

\item $a_8\ne 0$. \\ \medskip  Here we can solve (\ref{hcond1}) for $V_2(u_2)$ and 
substitute the result into (\ref{hcond2}) to obtain the equation
\begin{equation}\label{hcond5}\hbar^2\frac{d^3V_2}{du_2^3}-12V_2\frac{dV_2}{du_2}+12a_1\frac{dV_2}{du_2}=0, \end{equation}
where $a_1$ is an arbitrary constant.  Solutions of (\ref{hcond5}) are further subject to the requirement that a solution $U_2(u_2)$ of equations (\ref{hcond3}) and (\ref{hcond4}) exists. The general solution of
(\ref{hcond5}) is $V_2(u_2)=\hbar^2 \wp(u_2-u_{2,0};g_2,g_3)+a_1$, where $u_{2,0}$, $g_2$ and $g_3$ are arbitrary constants.

\end{enumerate}
\section{Classification of systems on the 2-sphere separating in spherical coordinates}\label{section5}
We use the coordinates (\ref{cansp}) with $x=\sin(\theta),\ y=\phi$ and list the systems that admit a 3rd order symmetry operator.
 They fall into 4 classes: 
 
 \begin{enumerate} \item Systems that are 2nd order superintegrable. 
 
  \bigskip
 
 These systems are all known and they are classical (all parameters in the potential are arbitrary).
  \item Systems that are neither 2nd or 3rd order superintegrable.
  
  \bigskip
  
  These are special cases of 2nd order superintegrable systems, except that they depend on $\hbar$, so they can't be scaled.
  
  \item Systems with algebraically dependent generators $A$ and $B$.
  \item Systems that are truly 3rd order superintegrable.
  \end{enumerate}

\subsection{Systems that are 2nd order superintegrable}
 
    \begin{equation}\label{system 27} V_1(x)=\frac{c_1\sqrt{1-x^2}}{x}+c_2,\quad V_2(y)=\frac{c_3}{\cos^2(y)}+\frac{c_4\sin(y)}{\cos^2(y)}.\end{equation}

 This is the 2nd  order superintegrable system $S_7$, \cite{KKMP2001}.
    \begin{equation}\label{system 1}V_1(x)=\frac{c_1}{1-x^2}+c_2,\ V_2(y)=\frac{c_3}{\cos^2(y)}+\frac{c_4}{\sin^2(y)}.\end{equation}
  This is the 2nd order superintegrable system $S_9$, \cite{KKMP2001}. It is characterized by the fact that all of the parameters $\alpha_j$ and $\beta_{jk}$ are zero.
  These are the only 2nd order 4-parameter superintegrable systems.
  
  \begin{equation}\label{system 30} V_1(x)=0,\quad V_2(y)=c\exp(-2iy).\end{equation}
This complex potential system admits a 3rd order symmetry and also a 1st order symmetry, so it is 2nd order superintegrable, the 2nd order system $S_5$, \cite{KKMP2001}.
 It is PT-symmetric so the energy eigenvalues are real.

  All other 2nd order systems are special cases of these. We will omit all classical special cases of these systems and include only purely quantum special cases.

\begin{enumerate}\item Quantum special cases.
  \begin{itemize}

 \item \begin{equation}\label{system 9} V_1(x)=\frac{\hbar^2}{c(1-x^2)}, \quad V_2(y)=0.\end{equation}
 This is a special case of $S_9$, \cite{KKMP2001},  but only quantum. The system admits a 1st order symmetry, so it is 2nd order superintegrable.
  \item \begin{equation}\label{system 18} V_1(x)=0,\quad V_2(y)=\frac{27 \hbar^2}{\cos^2(y)}.\end{equation}
 This system is  2nd order superintegrable and only quantum, a special case of $S_9$. It admits a 1st order symmetry.
\end{itemize}

\subsection{Systems that are  3rd order superintegrable but do not generate a cubic algebra}

   \begin{itemize}\item  \begin{equation}\label{system13}V_1(x)=0,\quad V_2(y)= \frac{-3\hbar^2(4\cos^4(y)-3)}{\cos^2(y)(2\cos^2(y)-3)^2}.\end{equation}
       This system is truly 3rd order superintegrable but only quantum.  The generators $\{H,A,B,C\}$ where $C=[A,B]$ do not close under commutation to
       form a cubic algebra. This is a special case of an isospectral deformation of the trigonometric Scarf potential, \cite{Quesne}. The bound state spectrum of $A$ is 
\[ \lambda = 2 \hbar^2(\frac12 + \nu) ^2 ,\quad \nu=0, 1, 2, \ldots,\] and the $A$ eigenfunctions are proportional to Jacobi exceptional orthogonal  polynomials of degree $\nu+2$ in $\sin(y)$.
\end{itemize}
 \end{enumerate}

\subsection{Systems with  algebraically dependent generators $A$ and $B$}
These systems are not 2nd order superintegrable and $C=[A,B]=0$. Thus the set $\{H,A,B\}$ is commutative so (by Burchnall-Chaundy theory), \cite{BurchnallChaundi}  there must be  a algebraic relation between $A$ and $B$ that we can compute. This permits us to exhibit an explicit first integral for the eigenfunctions of $A$. Thus these systems are special. We give the details for our first example. The other cases are similar.
    \begin{itemize}
     \item \begin{equation}\label{system 25} V_1(x)=0,\quad V_2(y)=\frac{4 \hbar^2}{\cos^2(2y)},\end{equation}
 Here $B$ is the formally self-adjoint operator
 \begin{equation}\label{B25} B=i\hbar^3\partial_y^3+8i\hbar^3\frac{2\cos^4(y)-2\cos^2(y)-1}{(2\cos(y)^2-1)^2}\partial_y-\frac{48i\hbar^3\sin(y)\cos(y)}{(2\cos(y)^2-1)^3}.\end{equation}
  The relationship is 
 \[  A^3-
 \frac18 B^2-4 \hbar^2 A^2 +4 \hbar^4 A =0.\] 
 \end{itemize}

  However, since $C=0$, the formally self-adjoint operators $B$ and $A$ admit  common eigenfunctions $g(y)$:
 \[ Bg=\mu g,\quad  Ag=\lambda g,\quad \lambda^3-\frac18\mu^2-4\hbar^2\lambda^2+4\hbar^4\lambda=0.\]
 Solving for $\mu$  we find
 \[ \mu=\pm 2\sqrt{2}(\lambda-2\hbar^2)\sqrt{\lambda}.\]
   Now consider the equation 
 \[ (Bg-\mu g)+a\frac{d}{dy}(Ag-\lambda g)=0 \]
for  any constant $a$. Choosing $a=2i\hbar$ we eliminate the 3rd derivative term in $y$ and obtain the 1st order differential equation
\begin{align} \label{A25} &-\frac{2i\hbar}{g(y)}\frac{d g(y)}{dy}= \\
&\frac{-8\cos^6(y)\mu+16i\cos(y)\sin(y)\hbar^3+12\cos^4(y)\mu-6\cos^2(y)\mu+\mu}{(2\cos^2(y)-1)(8\cos^4(y)\hbar^2-4\cos^4(y)\lambda-8\cos^2(y)\hbar^2+4\cos^2(y)\lambda-\lambda)}, \nonumber
\end{align}
into which we substitute the two possibilities for $\mu$. These are explicit first integrals for the 2nd order differential equations satisfied by the eigenfunctions of $A$. 

\medskip

 {\bf Another example}:
 \begin{itemize}
     \item \begin{equation}\label{system 15} V_1(x)=0,\quad V_2(y)=\frac{N(y)}{D(y)},\end{equation}
       where {\small
       \begin{eqnarray*} N(y)&=&\\
      &-& 4\hbar^2\left(2\sqrt{3}\sin(y)\cos^3(y)+2\cos^4(y)+18\sqrt{3}\cos(y)\sin(y)\right.\\ 
      && \left. -21\cos^2(y)-81\right),\\
        D(y)&=&\\
        &16&\cos^6(y)+72\sqrt{3}\sin(y)\cos^3(y)+48\cos^4(y)+108\sqrt{3}\cos(y)\sin(y)\\
        &-&207\cos^2(y)+243.
       \end{eqnarray*}}
  This system admits a 3rd order symmetry $B$, but then $C=0$. Thus $A$ and $B$ are algebraically dependent and the system is not 3rd order superintegrable. The algebraic relationship between $A$ and $B$ is
  \[A^3-\frac{1}{8}B^2-\hbar^2 A^2+\frac{i\sqrt{3}\hbar^3}{3} B +\frac{\hbar^4}{4} A+\frac23\hbar^6 =0.\]
  \end{itemize}
  {The general $C=0$ case}
  \begin{itemize}
     \item \begin{equation}\label{system31} V_1(x)\ {\rm arbitrary}, \
\hbar^2\bigg(-\frac{dV_2}{dy} \frac{d^4V_2}{dy^4}+\frac{d^2V_2}{dy^2}\frac{d^3V_2}{dy^3}\bigg)+12\bigg(\frac{dV_2}{dy}\bigg)^3 = 0.\end{equation}
 This differential equation admits the Weierstrass $\wp$-function as a solution. The system admits a 3rd order symmetry provided the additional condition
 \begin{equation} \label{system31a} \hbar^3\frac{d^3V_2}{dy^3}-12\hbar V_2\frac{dV_2}{dy}-4z_1\frac{dV_2}{dy}=0 \end{equation}
 is satisfied for some constant $z_1$. However, this condition is always satisfied since (\ref{system31a}) is a first integral of (\ref{system31}).  Equation (\ref{system31a}) implies $C=0$, so that $A$ and $B$ are algebraically dependent. Indeed, they obey the relation 
 \[ -B^2+A^3+\alpha A^2=0,\]
 where 
 \[-i B=\ \frac{3\hbar}{2}\frac{dV_2}{dy}-\hbar^3\partial_y^3+\left(3\hbar V_2+\alpha\hbar\right)\partial_y.\]
  Thus this general system is not 3rd order superintegrable.
 \end{itemize} 
 
There are also  elementary function solutions of the nonlinear defining equation, none of which lead to 3rd order superintegrability.
 In particular the following systems are solutions:
\begin{itemize} \item \begin{equation}\label{system 33} V_2(y)=\hbar^2\left(\frac{1}{\sin^2(y)}+\frac{1}{\cos^2(y)}\right),\,\end{equation}
 This can be considered as a 0-parameter potential  within $S_9$. It has a 2-parameter 3rd order symmetry, $\beta_{31},\alpha_3$,  but if $\beta_{31}\ne0$ the   system doesn't close to a cubic algebra. If $\beta_{31}=0,\alpha_3 \ne 0$ we find $C=0$. Then $A$ and $B$ are related by
 \[ A^3-\frac18 B^2-4 \hbar^2 A^2 +4 \hbar^4 A=0.\]
 \end{itemize}
 \begin{itemize}
 \item \begin{equation}\label{system 34} V_2(y)=\frac{4\hbar^2}{(2\sin^2(y)-1)^2}.\end{equation} This system admits a 3rd order symmetry but $C=0$ so $A$ and $B$ are algebraically dependent:
 \[ A^3-\frac18 B^2-4 \hbar^2 A^2 +4 \hbar^4 A =0.\]
 \item
 \begin{equation}\label{system35} V_2(y) = \frac{\hbar^2(4b^2+c^2)}{(2b\cos^2(y)\pm c\sin(y)\cos(y)-b)^2}, \end{equation}
 This can be obtained from the Weierstrass $\wp$-function solution.  
 The system admits a 3rd order symmetry but $C=0$ so $A$ and $B$ are algebraically dependent. The relationship between $A$ and $B$ is 
 \[ A^3-\frac18 B^2-4 \hbar^2 A^2 +i
\frac{48 b^3 \hbar^3+12 b c^2 \hbar^3}{c^3}
 B +4\hbar^4 A +288\frac{(4b^2+c^2)^2 b^2\hbar^6}{c^6}=0.\]
\end{itemize}

\subsection{Systems that are truly 3rd order superintegrable}

In addition to the nonlinear solution (\ref{nonlinearsoln}) we have: 
    \begin{itemize}
     \item   \begin{equation}\label{system 20} V_1(x)=0,\quad V_2(y)=\frac{c}{(4\cos^2(y)-3)^2\cos^2(y)}.\end{equation}
 
This system is truly 3rd order superintegrable.
 It satisfies a cubic algebra of the form. 
\begin{equation}\label{[A,C]} [A,C]=\alpha A^2+\beta\{A,B\} +\gamma A+ \delta B+ \epsilon,\end{equation}
\begin{equation}\label{[B,C]} [B,C]= \mu A^3+\nu A^2-\beta B^2-\alpha \{A,B\}+\xi A-\gamma B+\zeta,\end{equation}
where 
\[ \gamma=\gamma_0+\gamma_1 H,\quad \delta=\delta_0+\delta_1 H,\quad \epsilon=\epsilon_0+\epsilon_1 H+\epsilon_2 H^2,\]
\[ \nu=\nu_0+\nu_1 H,\quad \xi=\xi_0+\xi_1 H+\xi_2 H^2,\quad \zeta=\zeta_0+\zeta_1 H+\zeta_2 H^2+\zeta_3 H^3.\]
\end{itemize}

 The final closure relation is 
 \begin{eqnarray} C^2=
  &&\frac{9\hbar^2}{2}\{A,B^2\}+576\hbar^2 A^4+9\hbar^2 BAB -576\hbar^2 H^3A+\nonumber\\ &&1728\hbar^2H^2A^2-1728\hbar^2HA^3-\frac{243\hbar^2}{4}B^2+(16704\hbar^4-576c\hbar^2)A^3\nonumber\\
  &&+(10080\hbar^4-1728c\hbar^2)H^2A+(-26784\hbar^4+1728c\hbar^2)HA^2\nonumber\\
  &&+576c\hbar^2 H^3+(27792\hbar^6-8928c\hbar^4)A^2-2304c\hbar^4H^2\nonumber\\
  &&+(-17280\hbar^6+11232c\hbar^4)HA+1728c\hbar^6H+(2592\hbar^8-5760c\hbar^6)A.\nonumber
 \end{eqnarray}
 The possible spectra for $H$ can be computed from these algebraic relations.

\section{Systems with  algebraically dependent generators as degenerations of $\wp$-potentials}
\label{section6}
The systems with algebraically dependent generators $A$ and $B$ can all be obained as degenerations of $\wp$-potentials.
The Weierstrass $\wp$-function with invariants $g_2,g_3$ satisfies the 1st order, nonlinear differential equation
\begin{equation}\label{deq_1}
(\wp')^2=4\wp^3-g_2\wp-g_3.
\end{equation}
We can parameterize the invariants $g_2$ and $g_3$ as 
\begin{equation}
g_2=2(e_1^2+e_2^2+e_3^2),\quad g_3=4e_1e_2e_3,
\end{equation}
so that
\begin{equation}\label{deq_2}
(\wp')^2=4(\wp-e_1)(\wp-e_2)(\wp-e_2),\quad e_1+e_2+e_3=0.
\end{equation}
When two of the roots $e_i$ coincide, (\ref{deq_2}) can be integrated in terms of elementary functions. We take $e_1=e_2$ so that $e_3=-2e_1$. The general solution of 
\begin{equation}\label{deq_3}
(\wp')^2=4(\wp-e_1)^2(\wp+2e_1)
\end{equation}
is 
\begin{equation}\label{hyperbolic_degeneration}
\wp(z-z_0;12e_1^2,-8e_1^3)=3e_1\mathrm{csch}^2\big(\sqrt{3e_1}(z-z_0)\big)+e_1
\end{equation}
or, making the replacement $e_1\to-e_1$,
\begin{equation}\label{trig_degeneration}
\wp(z-z_0;12e_1^2,-8e_1^3)=3e_1\mathrm{csc}^2\big(\sqrt{3e_1}(z-z_0)\big)-e_1.
\end{equation}
Either of these expressions degenerates to a rational solution as $e_1\to 0$:
\begin{equation}\label{rational_degeneration}
\wp(z-z_0;0,0)=\frac{1}{(z-z_0)^2}.   
\end{equation}

\subsection{Potentials from third-order superintegrability}

On the $2$-sphere we have the potential
\begin{equation}
V_2(u_2)=\hbar^2\wp( u_2-u_{2,0};g_2,g_3)+a_1
\end{equation}
Only the trigonometric degeneration is relevant here:
\begin{equation}
V_2(u_2)=\hbar^2\kappa^2\mathrm{csc}^2(\kappa( u_2-u_{2,0}))+\frac{\kappa^2}3+a_1,
\end{equation}
for an arbitrary real parameter $\kappa$.

On the $2$-hyperboloid we have the potential
\begin{equation}
V_2(u_2)=\hbar^2\wp( u_2-u_{2,0};g_2,g_3)+a_1.
\end{equation}
There are three degenerate potentials:
\begin{enumerate}
    \item $V_2(u_2)=\hbar^2\kappa^2\mathrm{csch}^2(\kappa( u_2-u_{2,0}))+\frac{\kappa^2}3+a_1$
    \item $V_2(u_2)=\hbar^2\kappa^2\mathrm{csc}^2(\kappa( u_2-u_{2,0}))+\frac{\kappa^2}3+a_1$
    \item $V_2(u_2)=\hbar^2 (u_2-u_{2,0})^{-2}+a_1$,
\end{enumerate}
where $\kappa$ is an arbitrary complex parameter.

\section{Higher order FD-superintegrable systems}\label{section7}
    A standard superintegrable system    on an $n$-dimensional Riemannian manifold (real or complex) is a system that possesses $2n-1$ functionally independent constants of the motion in the classical case and $2n-1$ algebraically independent symmetry operators in the operator case.
    If a system possess $2n-1$ linearly independent symmetries but they are not functionally (or algebraically) independent, we will call it {\it Functionally Dependent - superintegrable}, (or FD-superintegrable for short).   (These are to be distinguished from Functionally Linearly Independent systems, \cite{BKM2020}, of which there is only one example in two dimensions)

 In this paper we have found a number of FD-superintegrable systems of a rather simple type: $C=0$. For these systems the operators $A$ and $B$ commute. Further, they are ordinary differential operators in the variable $y$ alone. The Abelian algebra generated by $\{H,A,B\}$ does have structure because $A$ and $B$ satisfy an algebraic equation. It is clear, moreover that the machinery constructed here will work almost unchanged on any manifold with metric of the form $ds^2= dx^2+F(x) dy^2$, not just those of constant curvature.

Consider the 1D Hamiltonian
\[A= -\frac{\hbar^2}{2} \frac{d^2}{dy^2}f(y)+V_0(y)f(y)\]
and the symmetry operator 
\[B=W_0(y)+W_1(y)\frac{d}{dy}+W_2(y)\frac{d^2}{dy^2}+W_3(y)\frac{d^3}{dy^3}.\]
We require that $[A,B]=0$ and that $B$ is linearly independent of $A$.
The solution, unique up to a constant factor, is
\begin{equation} B=\frac{d^3}{dy^3}-\frac{-b_1\hbar^2+3V_0(y)}{\hbar^2} \frac{d}{dy}-\frac{3}{2\hbar^2}\frac{dV_0(y)}{dy}, \end{equation}
subject to the condition
\begin{equation}\label{system31A} -4b_1\frac{dV_0}{dy}\hbar^2-\frac{d^3V_0}{dy^3}\hbar^2+12V_0\frac{dV_0}{dy}=0.\end{equation}
This can be identified with equation (\ref{system31a}), so the treatment of (\ref{system31}) carries over immediately to this 1D system.

To summarize, the differential equation 
\begin{equation}\hbar^2\bigg(-\frac{dV_2}{dy} \frac{d^4V_2}{dy^4}+\frac{d^2V_2}{dy^2}\frac{d^3V_2}{dy^3}\bigg)+12\bigg(\frac{dV_2}{dy}\bigg)^3 = 0. \label{4th_order_ode}\end{equation}
can be integrated in terms of the Weierstrass $\wp$-function. This equation passes the Painlev\'{e} test with movable double poles as the only singularities. We can find the general solution.
       
This construction can be extended to general $n$. We present an example that is pertinent to 5th order FD-superintegrability  for the 2-sphere. The 5th order operator that commutes with $A$ is
\begin{align}
 B=& \frac{d^5}{dy^5}+c_4\frac{d^4}{dy^4}-\frac{5V_0}{\hbar^2}\frac{d^3}{dy^3} 
 -\frac{(-2c_2\hbar^2+8c_4V_0+15\frac{dV_0}{dy})}{2\hbar^2}\frac{d^2}{dy^2}-\\
 &\frac{-4c_1\hbar^4+16c_4\frac{dV_0}{dy}\hbar^2+25\frac{d^2V_0}{dy^2}\hbar^2-30V_0^2}{4\hbar^4}\frac{d}{dy}-\nonumber\\
 &\frac{-8c_0\hbar^4+16c_4\frac{d^2V_0}{dy^2}\hbar^2+16c_2\hbar^2V_0+15\frac{d^3V_0}{dy^3}\hbar^2-32V_0^2c_4-60V_0\frac{dV_0}{dy}}{8\hbar^4},\nonumber
\end{align}
  subject to the condition 
  \begin{eqnarray} \label{5thordercond}16c_1\frac{dV_0}{dy}\hbar^4+\frac{d^5V_0}{dy^5}\hbar^4-20V_0\frac{d^3V_0}{dy^3}\hbar^2-40\frac{dV_0}{dy}\frac{d^2V_0}{dy^2}\hbar^2+120V_0^2\frac{dV_0}{dy}=0.   
  \end{eqnarray}
    Here, $A$ and $B$ are related by the expression
    \begin{eqnarray}\label{algrel} B^2+a_1A^5+a_2BA^2+a_3A^4+a_4BA+a_5A^3+a_6B+a_7A^2+a_8A=0,\end{eqnarray}
   where 
  \[ a_1=\frac{32}{\hbar^{10}},\ a_2=-\frac{8c_4}{\hbar^4},\ a_3=\frac{16c_4^2}{\hbar^8},\ a_4=\frac{4c_2}{\hbar^2},\]
  \[ a_5=\frac{c_1-c_2c_4}{\hbar^6},\ a_6=-2c_0,\ a_7=\frac{-k_1}{2\hbar^{10}},\]
and $V_0(y)$ satisfies the equation
\begin{eqnarray}\label{5thordercond2} 48V_0^5&&+(64c_1\hbar^4+80\hbar^2\frac{d^2V_0}{dy^2}V_0^3+(-192c_4c_0\hbar^6-96c_2^2\hbar^6- \\
&&120\hbar^2\frac{dV_0}{dy}^2-12k_1)V_0^2+(-20\frac{d^2V_0}{dy^2})^2\hbar^4+32\hbar^6c_1\frac{d^2V_0}{dy^2}) \nonumber\\
&&-96\hbar^8(a_8\hbar^2+4c_0c_2-2c_1^2))V_0+(-16\hbar^6c_1+4\hbar^4\frac{d^2V_0}{dy^2})\frac{dV_0}{dy}^2\nonumber\\
&&-2\hbar^2(16c_0c_4\hbar^6+8c_2^2\hbar^6+k_1)\frac{d^2V_0}{dy^2}+128c_0c_1c_4\hbar^{10}+64c_1c_2^2\hbar^{10}\nonumber\\
&&-64c_0^2\hbar^{10}+(\frac{d^3V_0}{dy^3})^2\hbar^6+8c_1\hbar^4k_1-a_{10}=0\nonumber.\end{eqnarray}
Here (\ref{5thordercond2}) is an integrated form of (\ref{5thordercond}). See \cite{BurchnallChaundi} for the theory behind these results.
 
However, since $[A,B]=0$, the  formal operators $B$ and $A$ admit common eigenfunctions $g(y)$:
 \[ Bg=\mu g,\quad  Ag=\lambda g,\]
and 
\begin{eqnarray}\label{algrel1}\mu^2+a_1\lambda^5+a_\mu\lambda^2+a_3\lambda^4+a_4\mu\lambda+a_5\lambda^3+a_6\mu+a_7\lambda^2+a_8\lambda=0,\end{eqnarray}
a quadratic equation for $\mu$ as a function of $\lambda$. Differentiating the equation $Ag=\lambda g=0$ repeatedly we can derive linear equations for $\frac{d^5 g}{dy^5}, \frac{d^4 g}{dy^4}, \frac{d^3 g}{dy^3}, \frac{d^2 g}{dy^2}$, as functions of $\frac{dg}{dy},g$. Substituting these results into the equation $Bg-\mu g=0$ we obtain a first order differential equation for $g(y)$. Thus this construction enables one to express the eigenfunctions of $A$ in terms of a single integral.

 The equation (\ref{5thordercond}) passes the Painleve test. It appears in the  Cosgrove list, \cite{Cosgrove2000} and is equivalent to
 \[ u^{(5)}=20 u u^{(3)}+40 u' u'' -120 u^2 u'+\alpha u'\]
 via the change of variables 
 \[ V_0(y)=u(x),\quad y=\hbar x,\quad c_1=-\frac{\hbar^5}{16}\alpha.\]
 This is Cosgrove’s equation Fif-III \cite[Eq. 2.71]{Cosgrove2000} with $\lambda=\kappa=0$.
The equation can be integrated once to
\[ u ^{(4)} = 20 u u '' + 10(u ' )^2 - 40u^ 3 + \alpha y + \beta.\]
This is Cosgrove's equation F-V \cite[Eq. 1.7]{Cosgrove2000} with $\kappa  = 0$.
 The general solution is in terms of genus two hyperelliptic functions, but there are particular solutions in terms of elliptic functions and degenerations. This Burchnal-Chaundy construction  occurs in many fields, see for example the connection with the KdV hierarchy, equation \cite[Eq. 2.11]{GRT}. Here we are pointing out its significance in superintegrability theory. 

\medskip 

For all systems with $C=0$ we can use this method to reduce the solution of the $A$ eigenvalue equation to a single quadrature. The method also extends to systems where $B$ is of arbitrarily high order, $A$ is 2nd order and $C=0$, and always yields stucture equations relating $A$ and $B$ and a solution for the eigenvalues of $A$ up to a single quadrature.

  \section{The 2-hyperboloid in horocyclic coordinates} \label{section8}
 
We classify 3rd order superintegrable systems on the 2-sheet hyperboloid $s_1^2-s_2^2-s_3^2=1$ with horocyclic coordinates. In this section, we set $\hbar=\sqrt{2}i$ for simplicity. Then, with horocyclic coordinates $u_1,u_2$, (\ref{horcoords}), a separable Hamiltonian takes the form
\begin{equation}
H=u_2^2\left(\partial_{u_1}^2+\partial_{u_2}^2\right)+u_2^2(V_1(u_1)+V_2(u_2)).
\end{equation}
and 
\begin{equation}
A=\partial_{u_1}^2+V_1(u_1)
\end{equation}

We obtain the following systems.
  
 \subsection{Systems that are 2nd order superintegrable}
 
 \begin{itemize}
 \item 
 \[ V_1(u)=c_1+c_2 u_1+4c_3 u_1^2,\quad V_2(v)=c_3 u_2^2 \]
 This is the 2nd order superintegrable system $S1$, \cite{KKMP2001}.
 \item 
  \[ V_1(u)=\frac{c_1}{u_1^2}+c_2 +c_3 u_1^2,\quad V_2(v)=c_3 u_2^2 \]
  This is the 2nd order superintegrable system $S2$, \cite{KKMP2001}.
 \end{itemize}
 
 \subsection{Systems with  algebraically dependent generators $A$ and $B$.}
 
 {\bf A truly 3rd order system, but no cubic algebra.}

\medskip
 
  Here, $V_2(v)=0$ and $V_1(u)$ satisfies the nonlinear equation 
\begin{eqnarray}
&&8V_1+4u_1V_1'+6cV_1V_1'+cV_1'''=0 \nonumber
\end{eqnarray}
It admits the first integral
\[ V_1''=-\frac{4V_1^3c^2+16V_1^2cu_1-c^2(V_1')^2+16u_1^2V_1-4V_1'c+2c_1c}{2c(cV_1+2u_1)}. \]
This equation does not satisfy the Painlev\'e property.
The structure equation is 
\begin{equation}
[A,B]=2A=C
\end{equation}
which gives no information about the eigenvalues of the Hamiltonian, but shows that $A$ is a raising operator for the eigenvalues of $B$. Here, 
\[ B = \frac34cV_1 +\left(  u_1 +\frac32 cV_1\right)\partial_{u_1} +  u_2\partial_{u_2}+c\partial_{u_1}^3.\]
 {

{\bf Systems with  $C=0$} 

\medskip

  There are formally the same functionally superintegrable solutions for the hyperboloid as for the sphere, except that now $V_2(u_2)$ is arbitrary, and the ranges of the variables $u_1,u_2$ differ from those of $x,y$.

 \subsection{Systems that are truly 3rd order superintegrable.} 

Here, we have not classified the  quantum special cases, i.e. the quantum $\hbar$-systems that are special cases of classical superintegrable systems. 
\medskip

 {\bf The TTW solution}:  

\medskip

The TTW method applied to the system $S7$ with horocyclic coordinates leads to an infinite family of higher order superintegrable systems, and exactly one  system  is 3rd order superintegrable: 
\begin{equation}
U(u)=U_0+U_1 u_1+U_2 u_1^2, \quad V(v)=U_2 u_2^4.
\end{equation}
Thus $A=\partial_{u_1u_1}^2+U_0+U_1 u_1+U_2 u_1^2$. 
We can verify that the Hamiltonian with this potential is third order superintegrable with
\begin{align}
B=&\frac{2u_2U_2}{U_1}\partial_{u_1 u_1}^2\partial_{u_2}+\frac{(2U_2u_1+U_1)}{U_1}\partial_{u_1}^3
-\left(\frac{2U_2^2u_1^2}{U_1}+2U_2u_1+\frac12 U_1\right)u_2\partial_{u_2}\\
+&\frac{3U_1^2u_1+(4U_2u_2^2+6U_2u^2-U_1u_1+2U _0)U_1+ 8u_1(U_2u_2^2+\frac12 U_2u_1^2+\frac12 U_0)U_2}{2U_1}\partial_{u_1} \nonumber \\
+&\left(\frac{2u_2^2U_2^2}{U_1}+\frac{2U_2^2u_1^2}{U_1}+2U_2u_1\right)\nonumber
\end{align}

\medskip

{\bf The structure equations}:

\medskip

Then, with $C=[A,B]$ we have
\begin{eqnarray}{[A,C]} &=&-16\ U_2 B-\frac{32\ U_2^2}{U_1} A+\frac{4U_2(4U_0U_2-3U_1^2)}{U_1},\nonumber\\
{ [B,C]} &=&a_1 A^3+a_2A^2+a_3HA+a_4B+a_5A+a_6H+a_7,\nonumber\end{eqnarray}

with 
\begin{eqnarray} a_1&&=\frac{ -32U_2^2}{U_1^2},\ a_2 = \frac{12U_2(4 U_0U_2-U_1^2)}{U_1^2},\ a_3 = \frac{-64U_2^3}{U_1^2},\  a_4 = \frac{32U_2^2}{U_1},\nonumber \\
 a_5&&= -\frac{(16U_0^2U_2^2-8U_0U_1^2U_2+U_1^4-160U_2^3)}{U_1^2 },\   a_6  = \frac{16U_2^2(4U_0U_2-U_1^2)}{U_1^2 }    \nonumber           \\
 a_7&&= \frac{-4U_2^2(20U_0U_2-9U_1^2)}{U_1^2}.\nonumber\end{eqnarray}     
 \begin{eqnarray}C^2&&=b_1\{A,B^2\} +b_2A^4+b_3BAB
 +b_{4}B^2+b_{5}A^3+b_{6}HA^2+b_{7}A^2\nonumber
 \\
 &&+b_{8}HA+b_{9}A+b_{10}H+b_{11}\{A,B\}+b_{12},  \nonumber
 \end{eqnarray} where we have omitted the constants $b_j$.

The eigenvalues of $A$ are of the form $\lambda=-4\sqrt{U_2}\ n+{\rm constant}$, linear in $n$.
Note that this system is St\"ackel equivalent to a 3rd order Euclidean space superintegrable system in Cartesian coordinates.

\medskip

{\bf The remaining 3rd order nonlinear superintegrable systems.}

\medskip

These are the  systems treated in (\ref{horocyclicnonlinear}).

\section{Discussion and conclusions} \label{section9} We have developed a new approach to the classification of  higher order superintegrable systems on any 2D manifold that admit a separation of variables, not just Euclidean space. As a check we have  reproduced the striking Tremblay and Winternitz result that for Euclidean space and polar coordinates one of the 3rd order superintegrable  potentials can be expressed in terms of Painlev\'e VI.  As new results we show that  Painlev\'e VI also appears for 3rd order superintegrable systems on the 2-sphere and the 2-hyperboloid,  this occurrence is common for all constant curvature spaces.  We  derived the possible separable systems on the 2-sphere in spherical coordinates and the two sheet 2-hyperboloid in horocyclic coordinates that admit a third order symmetry operator and provide information about their symmetry algebras. Our aim is  to develop a convenient tool that can be used study higher order superintegrability on spheres, hyperboloids and more general Riemannian spaces, and to compare results between spaces.

\section*{Acknowledgments}
We thank Pavel Winternitz for helpful discussions and Adrian Escobar for pointing out the 
relevance of the paper \cite{KKM2010} to classification of 3rd order superintegrable systems.
W.M. was partially supported by a grant from the Simons Foundation (\# 412351 to Willard Miller, Jr.). 
B.K.B.\ acknowledges support from the G\"oran Gustafsson Foundation and the European Research Council, Grant Agreement No. 682537. \ I.M. was supported by the Australian Research Council Discovery Grant DP160101376 and Future Fellowship FT180100099.

%


\begin{thebibliography}{6}
%
\bibitem{BMM2020}Berntson, B.K., Marquette, I.  and Miller, W. Jr.,    A new approach to analysis of 2D  higher order quantum superintegrable systems,  To appear in  the proceedings volume for the Symposium Quantum Theory and Symmetry XI July, 2019, {\it CRM Series on Mathematical Physics},  	arXiv:1909.08654 [math-ph]. (2019).

\bibitem{KKM2010}  Kalnins, E.G.,  Kress, J.M. and  Miller, W, Jr., Superintegrability and higher order integrals for quantum systems, {\it J. Phys. A: Math. Theor.} {\bf43} (2010) 265205.

\bibitem{TTW}
Tremblay, F., Turbiner, V.A. and Winternitz, P.,
An infinite family of solvable and integrable quantum
systems on a plane.
{\em J.~Phys.~A:~Math.~Theor.} {\bf 42} (2009) 242001. 

\bibitem{TW} Tremblay, F. and Winternitz, P., Third order superintegrable systems separating in polar coordinates, {\it J. Phys.}, {\bf A43} ,175206 (2010).
\bibitem{Bocher} Kalnins, E.G.,  Kress, J,M. and Miller, W.  Jr,.  Separation of variables and Superintegrability: The symmetry of solvable systems,  {\it Instititute of Physics, UK}, 2018, ISBN: 978-0-7503-1314-8,     e-book. 


\bibitem{GW} Gravel, S., and Winternitz, P., Superintegrable systems with third-order integrals in classical and quantum mechanics,
{\it J. Math. Phys.}, {\bf 43}, 5902-5912, (2002). 
\bibitem{MarW} Marquette, I. and Winternitz, P., Superintegrable systems with third order integrals of motion {\it J. Phys. A:
Math. Theor.},  {\bf 41}, 304031, (2008). 

\bibitem{MSW} Marquette, I., Sajedi, M., and Winternitz, P. Fourth order superintegrable systems separating in Cartesian coordinates I. Exotic quantum potentials, {\it J. Phys. A: Math. Theor.}, {\bf 50} 315201 (2017). 


\bibitem{KKMP2001}  Kalnins, E.G.,  Kress, J.M. Miller, W.  Jr and Pogosyan, G.S.,
Completeness of superintegrability in two-dimensional constant curvature spaces, {\it J. Phys. A: Math Gen. } {\bf 34},  4705--4720, (2001)

\bibitem{Gravel}  Gravel, S.,          Hamiltonians separable in Cartesian
coordinates and third-order integrals of motion. {\it J. Math Phys.}, {\bf 45}, 1003-019, 2004.
\bibitem{IMA}  Symmetries and Overdetermined Systems of Partial Differential
Equations. (Editors Michael Eastwood and Willard Miller, Jr), IMA Volume
144: 
  Springer, New York, 2007.
\bibitem{TTW2}  Tremblay, F.,   Turbiner, A.  and  Winternitz, P., Periodic orbits for an infinite family of classical superintegrable 
systems. {\it J. Phys. A: Math. Theor.} {\bf 43} (2010) 015202.  
\bibitem{JH}
Hietarinta, J.,
Direct  methods for the search of  the second invariant.
Physics Report, 1987, Vol.147, pp. 87-154.
\bibitem{CQ10}  Quesne, C., Superintegrability of the Tremblay-Turbiner-Winternitz quantum Hamiltonians on a plane for odd $k$.  {\it J. Phys. A: Math. Theor.} {\bf 43} 082001, 2010.
\bibitem{KKMPog07} Kalnins, E.J.,  Kress, J.M.  Miller, W.Jr., and Pogosyan, G.S., 
Infinite order symmetries for quantum separable systems. {\it Nuclei. nat.},  {\bf 68},  10, 1817-1824, 2007.

\bibitem{KKM2018}   Kalnins, E.G.,   Kress, J.M. and Miller, W. Jr.,  Separation of variables and Superintegrability: The symmetry of solvable systems, 
Instititute of Physics, UK,  2018, ISBN: 978-0-7503-1314-8, 
http://iopscience.iop.org/book/978-0-7503-1314-8.


\bibitem{KKM2005} Kalnins, E.G., Kress, J.M., and Miller W. Jr., Second order superintegrable systems in conformally flat spaces. 3. 3D classical structure theory, {\it J. Math. Phys.}, {\bf 46}, 103507, (2005)

\bibitem{BKM2020} Berntson, B.K., Kalnins, E.G. and Miller, W. Jr. Classification of Calogero-like 2nd order superintegrable systems in 3 dimensions. (submitted) arXiv:2004.00933 [math-ph]. (2020).

\bibitem{BurchnallChaundi} Burchnall, J.L. and  Chaundy, T.W., Commutative ordinary differential
operators, {\it Proc. London Math. Soc. Ser. 2}, {\bf 21}. (1923) 420-440; {\it Proc.
Royal Soc. London Ser. A}, {\bf 118} (1928) 557–583.
\bibitem{Quesne} Quesne, C., Solvable rational potentials and exceptional orthogonal polynomials in supersymmetric quantum mechanics,  {\it Symmetry, Integrability and Geometry: Methods and Applications}. SIGMA 5 (2009), 084, 24 pages.
\bibitem{Cosgrove2000}Cosgrove, C.M., Higher-order Painlev\'e equations in the polynomial class I: Bureau
symbol P2, {\it  Stud. Appl. Math}. {\bf 104}, 1 (2000)

\bibitem{Cosgrove2000b} Cosgrove, C.M., Chazy Classes IX–XI Of Third‐Order Differential Equations, {\it Stud. Appl. Math}. {\bf 104} 171 (2000). 

\bibitem{Cosgrove1993}  Cosgrove, C.M. and  Scoufis, G., Painlev\'e Classification of a Class of Differential Equations of the Second Order and Second Degree, {\it Stud. Appl. Math.} {\bf 88} 25 (1993). 

\bibitem{GRT} Gesztesy, F., Ratnaseelan, R. and Teschl, G., 
The KdV Hierarchy and Associated Trace Fomulas,  Recent Developments in Operator Theory and its Applications, I., Gohberg (ed.) et al., 125–163, {\it Oper. Theory Adv. Appl.}, {\bf 87}, Birkh\"auser, Basel, 1996.
\end{thebibliography}
\end{document}